\documentclass[aps,pra,twocolumn,eqsecnum,nofootinbib,longbibliography,superscriptaddress]{revtex4-2}
\usepackage{amsmath}
\usepackage{graphicx}
\usepackage{mathrsfs}
\usepackage{color}
\usepackage{hyperref}
\usepackage{lipsum}
\usepackage[procnames]{listings}
\usepackage{layouts}
\usepackage{epstopdf}
\usepackage{dsfont}
\usepackage{bbm}
\usepackage{physics}

\newcommand{\wthreej}[6]{\begin{pmatrix}
  #1 & #2 & #3 \\
  #4 & #5 & #6 
\end{pmatrix}}

\begin{document}

\title{Systematic design of a robust half-W1 photonic crystal waveguide\\ for interfacing slow light and trapped cold atoms}

\author{Adrien Bouscal}
\affiliation{Laboratoire Kastler Brossel, Sorbonne Universit\'e, CNRS, ENS-Universit\'e PSL, Coll\`{e}ge de France, 4 place Jussieu, 75005 Paris, France} 

\author{Malik Kemiche}
\affiliation{Centre de Nanosciences et de Nanotechnologies, CNRS, Universit\'e Paris-Saclay, 91120 Palaiseau, France}
\affiliation{IMEP-LAHC, Univ. Grenoble Alpes, Univ. Savoie Mont Blanc, CNRS, Grenoble INP, 38000 Grenoble, France}

\author{Sukanya Mahapatra}
\affiliation{Centre de Nanosciences et de Nanotechnologies, CNRS, Universit\'e Paris-Saclay, 91120 Palaiseau, France}

\author{Nikos Fayard}
\affiliation{Université Paris-Saclay, Institut d'Optique Graduate School, CNRS, Laboratoire Charles Fabry, 91127 Palaiseau, France}

\author{J\'er\'emy Berroir}
\affiliation{Laboratoire Kastler Brossel, Sorbonne Universit\'e, CNRS, ENS-Universit\'e PSL, Coll\`{e}ge de France, 4 place Jussieu, 75005 Paris, France} 

\author{Tridib Ray}
\affiliation{Laboratoire Kastler Brossel, Sorbonne Universit\'e, CNRS, ENS-Universit\'e PSL, Coll\`{e}ge de France, 4 place Jussieu, 75005 Paris, France} 

\author{Jean-Jacques Greffet}
\affiliation{Université Paris-Saclay, Institut d'Optique Graduate School, CNRS, Laboratoire Charles Fabry, 91127 Palaiseau, France}

\author{Fabrice Raineri} 
\author{Ariel Levenson} 
\author{Kamel Bencheikh}
\affiliation{Centre de Nanosciences et de Nanotechnologies, CNRS, Universit\'e Paris-Saclay, 91120 Palaiseau, France} 

\author{Christophe Sauvan} 
\affiliation{Université Paris-Saclay, Institut d'Optique Graduate School, CNRS, Laboratoire Charles Fabry, 91127 Palaiseau, France}

\author{Alban Urvoy}
\email[Corresponding author: ]{alban.urvoy@sorbonne-universite.fr}
\author{Julien Laurat}
\affiliation{Laboratoire Kastler Brossel, Sorbonne Universit\'e, CNRS, ENS-Universit\'e PSL, Coll\`{e}ge de France, 4 place Jussieu, 75005 Paris, France} 

\date{\today}
\begin{abstract}
Novel platforms interfacing trapped cold atoms and guided light in nanoscale waveguides are a promising route to achieve a regime of strong coupling between light and atoms in single pass, with applications to quantum non-linear optics and quantum simulation. A strong challenge for the experimental development of this emerging waveguide-QED field of research is to combine facilitated optical access for atom transport, atom trapping via guided modes and robustness to inherent nanofabrication imperfections. In this endeavor, here we propose to interface Rubidium atoms with a photonic-crystal waveguide based on a large-index GaInP slab. With a specifically tailored half-W1 design, we show that a large chiral coupling to the waveguide can be obtained and guided modes can be used to form two-color dipole traps for atoms at 116~nm from the edge of the structure. This optimized device should greatly improve the level of experimental control and facilitate the atom integration.
\end{abstract}
\maketitle

\section{Introduction}

Interfacing cold neutral atoms and photons guided in nanoscale waveguides has raised a large interest over the recent years, with a wealth of emerging opportunities~\cite{Dibyendu2017, Chang2018, Sheremet2023}. Arrays of atoms can be trapped in the evanescent field of guided modes and the strong transverse confinement enables to increase the individual atom-photon coupling in single pass. Remarkable experimental advances have been obtained with optical nanofibers~\cite{Vetsch2010, Goban2012,Nieddu2016,Solano2017}, exploiting collective effects and chiral properties to realize various all-fibered functionalities \cite{Gouraud2015, Corzo2016, Sorensen2016, Corzo2019, Mitsch2014, Prasad2020}. Beyond nanofibers, tailored dispersion relations that can be obtained in photonic-crystal waveguides (PCW) offer unique features~\cite{Chang2018}. While the atom-photon coupling can be strongly enhanced near a band edge, where guided modes can propagate slowly, atom-photon bound states can also appear for an atomic transition within a band gap, with the capability to implement tunable long-range atom-atom interactions. These features led to a variety of theoretical proposals for applications in quantum optics and many-body physics~\cite{Douglas2015, GonzalezTudela2015, Bello2022}.

Despite the promises of this new waveguide-QED paradigm, trapping atoms in the vicinity of photonic-crystal waveguides is still at its infancy. This combination is a daunting task due to stringent requirements when considering real physical implementations. A first challenge is to keep the atoms as static as possible close to the structure, so that they can interact with the evanescent mode. While tweezers can be used to maintain the atoms at a fixed distance \cite{Thompson2013, Kim2019, Liu2022, Will2021}, it is challenging to make an array of such atoms at distances on the 100 nm range. Dipole trapping by the evanescent field of guided modes is necessary but it has remained an important roadblock. Up to now, only a corrugated slot waveguide (so-called alligator waveguide) \cite{Goban2014,Goban2015,Hood2016,Burgers2019,Yu2014} has been implemented and first pioneering demonstrations obtained, albeit with a limited number of atoms and without stable trapping in the evanescent field. Some theoretical proposals on novel interesting structures supporting atom trapping in the evanescent field have emerged since, such as a slot \cite{Yu2017} or a comb waveguide~\cite{Fayard2022}. Structures must also provide a large optical access to bring atoms close to their surface. Eventually, in order to push experimental development, great care should be put in ensuring that the structure is robust against fabrication imperfections.

\begin{figure*}[t]
  \includegraphics[width = 0.92\textwidth]{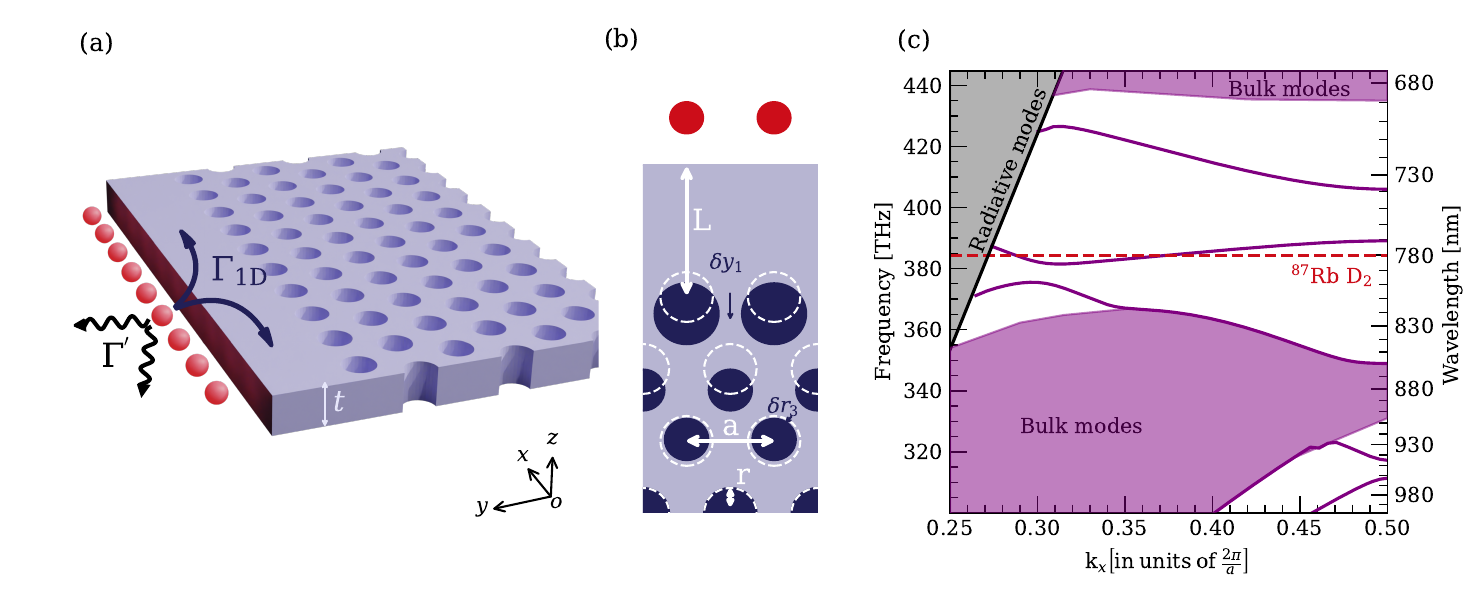}
  \caption{A half-W1 slow-mode photonic crystal waveguide coupled to cold atoms. (a)~Sketch of the waveguide with an array of $^{87}$Rb atoms trapped in the proximity, along the edge. $\Gamma_{\mathrm{1D}}$ and $\Gamma'$ correspond to the decay rates in the guided mode and in the radiation continuum, respectively. The structure is etched in a GaInP membrane (refractive index $n = 3.35$) suspended in air, with a slab thickness $t$ of 150~nm. (b)~2D scheme of the optimized photonic-crystal geometry. The initial unshifted and regularly distributed holes are shown as white dashed lines. For the first three rows the position of the holes can be shifted along $y$ and their radius tuned, amounting to 6 parameters ($\delta y_i$,$\delta r_i$), $i \in \{1,2,3\}$. For the sake of clarity, only two parameters ($\delta y_1$ and $\delta r_3$) are displayed. (c)~Bandstructure of the optimized structure calculated via FDTD simulation. We see a band gap for the selected polarization (transverse-electric) between 365 and 435~THz. The bulk modes propagate within the slab but are not guided on the edge of the PCW while the radiative modes are not guided at all. The $^{87}$Rb D$_2$ line transition frequency is aligned with the linear part of a guided band, defined as the $\textit{slow mode}$ in the text.}
  \label{fig:structure}
\end{figure*}

In this paper, we design a novel platform for interfacing trapped cold atoms and a slow-mode photonic crystal waveguide. Building on the promises of W1 waveguides, made of a linear defect in a 2D photonic crystal, and initial work in~\cite{Zang2016}, we propose a tailored platform for trapping arrays of Rubidium atoms in its proximity, as sketched in figure~\ref{fig:structure}(a). Waveguides based on a 2D photonic crystal etched in a large refractive-index slab have well-known strengths and are widely used in the telecom range. Many techniques have been developed to shape their dispersion curve with astounding precision~\cite{Petrov2004, Li2008, Frandsen2006, Colman2012, Vyas2022}. Strong coupling between guided light and a single emitter embedded in a W1 waveguide has been demonstrated~\cite{Arcari2014}, and successfully exploited for quantum operations~\cite{LeJeannic2022}. The proposed platform, sketched in figure~\ref{fig:structure}(a), can be seen as half a W1 waveguide, with the horizontal guidance mechanism relying on both photonic band gap and total internal reflection. As its W1 counterpart, it enables dispersion engineering but in addition offers a 2$\pi$ solid-angle optical access to the edge of the structure, allowing for simpler transport of atoms close to it~\cite{Zang2016}. We use a large refractive index GaInP slab that facilitates the design by offering more flexibility in the engineering of guided modes and band gaps, and we show how to trap atoms in the proximity via the evanescent field of additional guided modes. 
Our effort focuses at each step on making the design robust to imperfections and on assessing the experimental feasibility of the full platform. 

Interestingly, this tailor-designed platform exhibits high chiral coupling. This feature is crucial, as introducing chirality in 1D chains of emitters can modify its properties and give rise to new physics~\cite{Fedorovich2022}. From a technical point of view, chirality makes some theoretical models, such as the MPS formalism, more tractable. This allows for simulating bigger systems more accurately, bridging the gap for better agreements between simulations and experimental data~\cite{Mahmoodian2020}. Finally, chiral coupling can be a tool for generating non-Gaussian states of light~\cite{Kleinbeck2023}, highly entangled atomic states~\cite{Buonaiuto2019} and to implement novel quantum information protocols~\cite{Lodahl2017, Li2018}. To the best of our knowledge, our proposed platform is the only one combining chirality and large individual coupling with an array of atoms. \\

This paper is organized as follows. First, in section~\ref{sec:Purcell} we present the specific platform based on a half-W1 waveguide realized in a GaInP slab with a high refractive index. We detail the optimization of the dispersion curve and the resulting robustness to nanofabrication imperfections, and then provide the achievable atom-photon coupling. Second, in section~\ref{sec:trap} we show that guided modes can be used to trap atoms in the proximity of the waveguide via a two-color evanescent dipole trap. Stable traps down to 116~nm from the surface are obtained with low powers that are compatible with nanophotonic systems. A summary and outlook are provided in section~\ref{sec:conclu}.

\section{Engineered half-W1 waveguide for Rubidium atoms}\label{sec:Purcell}

In this section we introduce the specific half-W1 slow-mode waveguide designed in this work, based on GaInP. We identify the required geometrical parameters and then present the optimizations performed to increase the robustness to fabrication imperfections, leading thereby to linear bands. Finally, the expected coupling to the guided mode for atoms in the proximity of the surface is computed. This coupling will be expressed in terms of the Purcell factor $\Gamma_{\mathrm{1D}}/\Gamma_0$, where $\Gamma_{\mathrm{1D}}$ is the decay rate of a single atom in the waveguide mode and $\Gamma_0$ the decay rate of a single atom in vacuum.

\subsection{Description of the half-W1 GaInP waveguide}

A periodic modulation of the refractive index has deep consequences on light propagation. It enables a coupling between forward and backward propagating waves, opening photonic band gaps where light propagation is forbidden within a given frequency range. At the edge of these band gaps, the group velocity vanishes \cite{Joannopoulos2007} and the Purcell factor diverges (see Appendix \ref{appendixA}). 

Motivated by the proposal in \cite{Zang2016}, we study a similar structure with a different material: GaInP. This material has been chosen for its advantageous optical and electronic properties. GaInP has a wide electronic band gap below $1.85$~eV \cite{Schubert2005}, and as such is transparent for a wide range of wavelengths (from 670~nm up), meaning it could be used with several alkali. At 780~nm, its refractive index is $n = 3.35$, reaching $3.55$ at the electronic band edge. 
This large index contrast with the air gives rise to band gaps that are wider and further away from the light line~\cite{Joannopoulos2007}, allowing for more flexibility in the design of the dispersion curves (both for the slow and the trapping modes). 
Finally, this material has attracted some attention in recent years as it is very convenient to operate in the telecom band due to its low two-photon absorption \cite{Combrie2009}, and growth and fabrication processes have therefore been developed and well mastered.

As shown in figure~\ref{fig:structure}(b), the holes etched in the GaInP slab do not go up to the edge, leaving a few hundreds of nanometers of unperturbed slab where the light can propagate. 
Being based on a 2D slab rather than a 1D structure, this geometry should be quite rigid and prevent detrimental effects from low frequency mechanical modes.
The introduced symmetry breaking in the transverse direction allows for a more precise control on the dispersion properties of the waveguide since it offers extra degrees of freedom \cite{Lu2010}, while significantly improving the optical access. Transverse asymmetry has been harnessed in \cite{Nguyen2018} to create exotic dispersion bands such as Dirac cones, multivalleys, or flat bands. Arrays of $^{87}$Rb will then be trapped near the edge of the waveguide thanks to a two-color dipole trap, at 116~nm from the surface. For comparison, in tapered nanofiber platforms, atoms sit more than 200~nm away from the silica fiber.

\begin{figure}[t]
  \includegraphics[width = 0.49\textwidth]{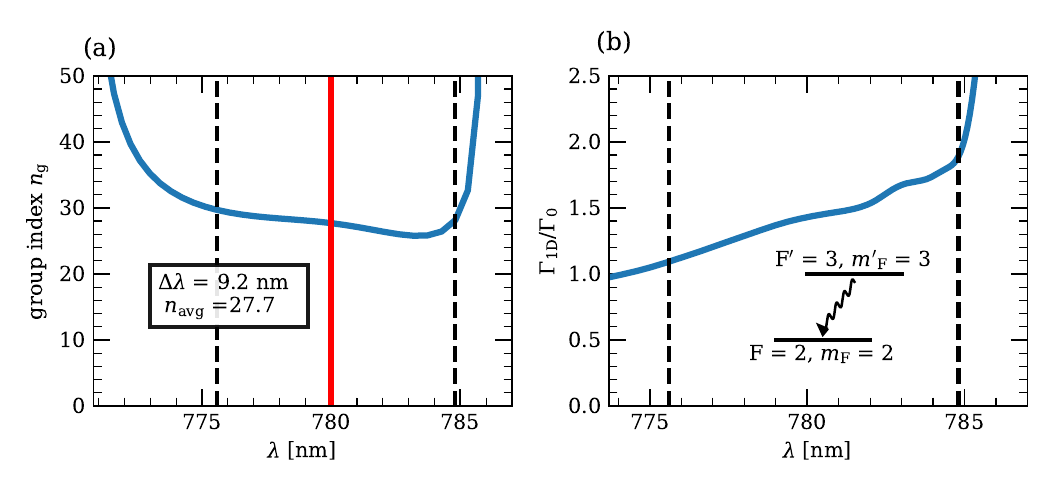}
  \caption{Dispersion and atom-coupling properties for the half-W1 waveguide with the structure optimization specified in Table \ref{table:shifts}. (a)~Calculated group index $n_\mathrm{g}$ for the slow mode. The dotted lines delimit the linear band region where the group index value is constant up to 15\%. (b)~Calculated Purcell factor $\Gamma_{\mathrm{1D}}/ \Gamma_0$ over the same range, for atoms trapped at 116 nm from the structure on the cyclic transition of the D$_2$ line. As it can be seen, $n_\mathrm{g}$ is not the only parameter affecting this ratio, i.e., the field structure is also changing, but it is still critical as it diverges with $n_\mathrm{g}$ just outside the plateau.}
  \label{fig:groupindex}
  \end{figure}

For a given thickness $t$, chosen here to be $t=150$~nm, the first step to determine the geometrical parameters consists in finding the lattice period $a$ and hole radius $r$ of the bulk 2D photonic crystal that allow for a band gap at the $^{87}$Rb D$_2$ transition. Indeed the width and position of the band gap is entirely determined by these values \cite{Joannopoulos2007}. The band gap has to be wide enough to allow for at least two guided modes, one that crosses 780~nm, and a blue-detuned one for trapping, as described later. Guided bands appear when introducing the defect at the edge, and we can align the band of interest with respect to the D$_2$ line by adjusting the width $L$.

Given these constraints, the geometrical parameters of the waveguide are found to be: $a$ = 212~nm, $r$ = 63~nm and $L$ = 337~nm for $t$~=~150~nm. The corresponding band structure, computed with the 3D FDTD software \textit{Lumerical}\footnote{2022 Ansys Lumerical simulation software based on the finite-difference time-domain (FDTD) method. \url{https://www.lumerical.com/}} is displayed in figure~\ref{fig:structure}(c) (represented over the $1^{\mathrm{st}}$ Brillouin zone, by application of Bloch's theorem). Three guided bands can be found inside the band gap of the 2D photonic crystal between 365 and 435~THz. The bulk modes are guided in the slab ($k_z$ imaginary) but can propagate in any direction in the plane, even inside the 2D array of holes ($k_x, k_y$ real). Above the light cone, radiative modes have a real $k$ vector in all directions and are therefore not guided.

\subsection{Imperfection-robust band engineering}

\begin{figure*}[t!]
  \includegraphics[width = 0.81\textwidth]{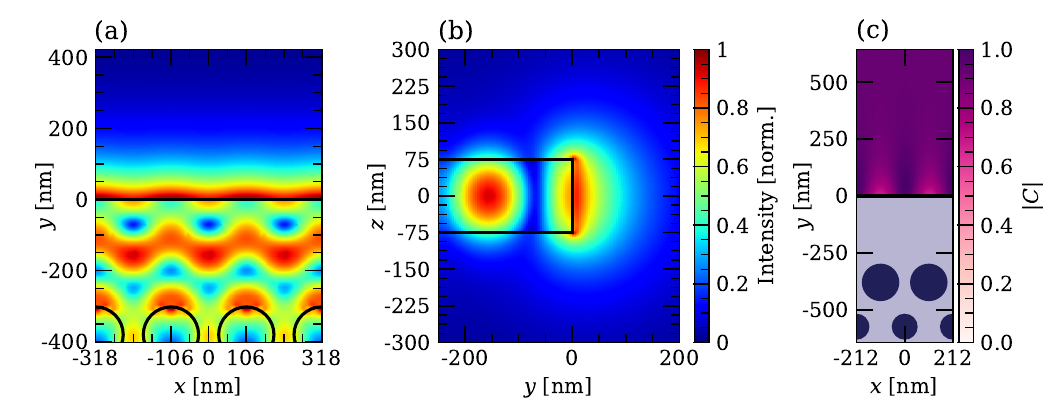}
  \caption{Structure of the forward-propagating slow mode at the $^{87}$Rb D$_2$ line frequency. (a) Normalized intensity, in the $(x,y)$-plane at $z$ = 0. (b) Same in the $(y,z)$-plane at $x = -a/2$, i.e., crossing the hole nearest to the slab edge. The mode is strongly expelled into the vacuum around the edge of the waveguide. (c) Polarization ellipticity $z$-component $C_z$ in the ($x$,$y$) plane at $z$ = 0. The other components of the ellipticity vector are 0. $|C_z| = 0$ indicates a linear polarization, while we have $|C_z|$= 1 for a circularly polarized light. Close to the edge, the polarization has a large circular component due to the strong longitudinal component that appears when light is confined at the nanoscale. By taking $z$ as the quantification axis, the polarization will be close to $\sigma^+$ for atoms trapped in the proximity ($92$ to $99\%$ fraction at 116 nm from the surface).}
 \label{fig:mode}
 \end{figure*}

Nanofabrication inherently leads to imperfections, even if errors below 2~nm can be reached~\cite{Asano2006}. A specific effort has been put in our design process to minimize the impact of such imperfections, thereby facilitating an experimental realization.

As the Purcell factor diverges at a band edge, one naive approach could be to align the D$_2$ line frequency $\omega_\mathrm{a}$ to any band edge of the band structure. However, fabrication imperfections, to first order, lead to a shift of the energy of the band \cite{Faggiani2016}. The flatter the band, i.e., the smaller the group velocity, the more it is vulnerable to a shift in frequency \cite{Soljacic2004}. If the D$_2$ line is aligned with the band edge, an infinitesimal shift to a lower energy will bring the atomic transition in the band gap of the 2D photonic crystal, impeding the propagation of the emitted light. In addition, the disorder in the frequency shift along the waveguide (from the randomness of the imperfections) can lead to strong localization of light inside the crystal~\cite{Mazoyer2009,Patterson2009,Faggiani2016}.

Following \cite{Zang2016}, two main criteria are to be considered when assessing the robustness of a structure: the group velocity has to be independent of frequency around the atomic transition, i.e. $\partial v_\mathrm{g}/\partial\omega\rvert_{\omega_\mathrm{a}} \sim 0 $, and the distance of the atomic transition to the band edge $\Delta \omega = |\omega_\mathrm{a} - \omega_\mathrm{e}|$ has to be as large as possible.

Designing slow modes with \textit{linear bands} (i.e., an almost constant, large group index $n_g$ over the widest range of $\omega$ possible) allows us to fulfill these two criteria. First, a linear dispersion corresponds to a vanishing group velocity dispersion (GVD) and the atom-photon coupling is proportional to the group index. Moreover, as shown in~\cite{Li2008}, it is possible to design a slow and linear band over a wide spectral range. As most fabrication imperfections lead to a shift $\Delta \omega$ of the guided bands, both these constraints aim at placing the relevant frequency at a position on the band where a small shift will affect the dispersion at the given frequency only slightly. It has been shown that \textit{linear bands} can be achieved in at least two types of asymmetric PCWs \cite{Nguyen2018}. Achieving such vanishing group velocity dispersion has been extensively studied in the context of W1 waveguides, by tuning the position of rows of holes \cite{Li2008, Wu2010,Liang2011}, chirping the waveguide properties \cite{Mori2005}, or changing the size of the holes \cite{Frandsen2006}. 

We note that such engineering is only possible for non-fundamental modes, which have a faster decay outside the structure~\cite{Krauss2007}. These modes come with non-trivial impedance and mode matching, which have been efficiently addressed in the literature using tapered regions~\cite{Hugonin2007, Hamel2013}.

\begin{table}[!b]
\centering
\begin{tabular}{ |p{1.2cm}||r|r| }
 \hline
 Row& Position $\delta y$ (nm) & Radius $\delta r$ (nm)\\
 \hline
 1   & $+42.7$  & $+14.2$\\
 \hline
 2 &   $+53.8$ &  $-11.2$  \\
 \hline
 3 & $-3.7$ & $-10.8$\\
 \hline
\end{tabular}
\caption{Optimal changes in row positions and holes radii via automatic differentiation optimization. All the rows after the third one are unperturbed.}
\label{table:shifts}
\end{table}

Inspired by these previous strategies, we optimize the shape of the slow-mode band by tuning the geometry of the 2D photonic crystal. As depicted in figure~\ref{fig:structure}(b), the 6 independent optimization parameters are the radius of the first three rows of holes as well as their position along the $y$ axis ($\delta r_i$, $\delta y_i$), $i \in \{1,2,3\}$. As full 3D FDTD simulations are computationally intensive, we use the approximate method of Guided Mode Expansion (GME) \cite{Andreani2006} thanks to the \texttt{legume} solver \cite{Minkov2020} to compute faster the shape of the guided band. At each iteration, a cost function enforcing the minimization of the group velocity dispersion (averaged over the wave vector interval) while setting a target $n_\mathrm{g}$ value is evaluated and the ($\delta r_i$, $\delta y_i$) varied thanks to automatic differentiation. After a few hundred iterations we obtain the optimal shifts for achieving this target $n_\mathrm{g}$ value over the widest possible spectral range. Finally, the optimized structure is simulated in full 3D FDTD to validate the results from the approximate GME method.

  \begin{figure*}[t!]
  \includegraphics[width = 0.81\textwidth]{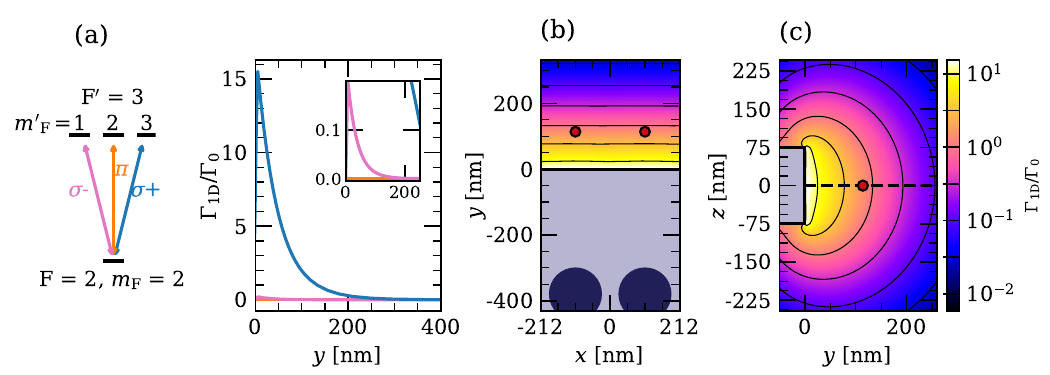}
  \caption{Excitation rates for $^{87}$Rb atoms in the waveguide proximity. (a) Allowed transitions on the D$_2$ line for an atom in $|\mathrm{F}=2, m_\mathrm{F} = + 2\rangle$. Because of the large $\sigma^+$ component ($\sim 91\%$ at the position of the atoms) and the values of the Clebsch-Gordan coefficients, the excitation probability to the $|\mathrm{F'}=3, m_{\mathrm{F'}} = + 3\rangle$ is 100 times higher than the $\sigma^-$ channel. The inset provides a zoom. (b) Purcell factor in the XY plane, at $z=0$. (c). Purcell factor in the YZ plane, at $x=-a/2$. The red dots indicate the position of the atoms at 115 nm from the surface.}
  \label{fig:G1D}
\end{figure*}

In order for this optimization to give relevant results, $n_\mathrm{g}$ has to be set to an experimentally realistic value, ideally below 60. Indeed, experiments have shown that it is extremely challenging to reach higher values for the group index without losses \cite{Mazoyer2010}. The most concluding optimization results are obtained for a target around $n_\mathrm{g} = 30$. The shifts in position and radius after optimization are given in table~\ref{table:shifts} and the corresponding band structure is presented in figure~\ref{fig:structure}(c). Figure \ref{fig:groupindex}(a) shows that we engineered a band with a constant group index of 28 over a 9~nm range, and hence reach similar performance than a previous optimization of a W1 waveguide~\cite{Li2008}.
This feature offers a two-fold advantage. In addition to making it robust to shifts caused by fabrication imperfections, the optimization enables using the half-W1 waveguide in a large bandwidth regime ($\geq 4$~THz) with very little dispersion. 

Finally, since the Purcell factor is proportional to the group index $n_\mathrm{g}$ (see Appendix \ref{appendixA}), we obtain with our optimization strategy an almost constant Purcell factor in this frequency range as seen in figure~\ref{fig:groupindex}(b). The residual variation arises from small changes in the structure of the electric field when moving along the guided band.

\subsection{Strong chiral coupling to the slow mode}

Given the optimized design, we now turn to the interaction between the slow mode and the $^{87}$Rb atoms in the vicinity of the waveguide. Taking into account the multilevel character of Rubidium, we defined a transition-dependent Purcell factor in equation~\eqref{eq:Gamma1D_multilevel} given in \ref{appendixA}. The group velocity $v_\mathrm{g}$ is evaluated from the simulated band structure, while the other terms are computed from the field map of the guided mode (as the Purcell factor is proportional to the slow-mode intensity). Figures~\ref{fig:mode}(a-b) show that in order to have the maximum coupling, the atoms should be trapped close to the edge of the waveguide, aligning them to the holes of the first row.

Figure~\ref{fig:mode}(c) shows that the guided mode has a strong circular polarization along the whole structure quantified by the ellipticity vector $\mathbf{C} = \Im\left[(\mathbf{E} \cross \mathbf{E}^*)/|\mathbf{E}|^2\right]$. This is reminiscent of the polarization of the light around nanofibers and comes from the longitudinal component of the electric field that appears when light is strongly confined. 
We note that this is also allowed by the slow-mode being far from the edge of the Brillouin zone, preventing time-reversal symmetry to cancel out the circularly polarized component of the slow mode~\cite{Mahmoodian2017}.
By choosing a quantization axis perpendicular to our waveguide (along $z$), the polarization of the forward propagating mode seen by the atoms is predominantly $\sigma^+$. The selectivity is further enhanced by the Clebsch-Gordan coefficients which favor transitions at larger $m_F$. This becomes explicit in figure~\ref{fig:G1D}(a) which shows the excitation rate $\gamma_{\mathrm{exc}}^+/\Gamma_0$ of an atom in the state \mbox{$|F=2, m_F = + 2\rangle$} with a resonant guided mode propagating along increasing $x$. Decay and excitation rates are distinct as atoms can be excited with a mode propagating in either direction while decay happens in both, see \ref{appendixA}. The $\sigma^+$ transition is stronger than the $\sigma^-$ transition by two orders of magnitude, while the $\pi$ transition is completely suppressed as the mode does not have any $E_z$ component in the symmetry plane. The excitation of an atom by a guided mode is thus highly chiral. 
Once in \mbox{$|F'=3, m_{F'} = + 3\rangle$}, guided mode emission will occur preferably in forward propagation, with a probability given by the polarization fraction, $92 \%$ at the position of the atoms.

Besides, figure~\ref{fig:G1D}(a) shows that for atoms in state \mbox{$|F=2, m_F = + 2\rangle$} and trapped at 116~nm from the edge, the Purcell factor reaches a value of 0.71. As shown in figures~\ref{fig:G1D}(b) and \ref{fig:G1D}(c), a small modulation in the $x$ direction exists and the value of the Purcell factor decays rapidly as a function of the distance to the surface. To quantify the coupling of the atoms to the guided mode we also define the $\beta$ factor, $\beta = \Gamma_{\mathrm{1D}}/\Gamma_{\mathrm{tot}}$ with $\Gamma_{\mathrm{tot}} = \Gamma_{\mathrm{1D}} + \Gamma'$, and $\Gamma'$ the decay rate in all the radiation modes other than the guided slow mode. Because of the complex shape of the local density of states accessible to the atoms, the behaviour of $\Gamma'$ is hard to infer. We performed a numerical calculation of $\Gamma'$ using the same method as in \cite{Fayard2022} and found that $\Gamma' \simeq 0.8 \Gamma_0$, at the position of the trap minimum, i.e at 116 nm from the surface.
We thus obtain $\beta = 0.47$, very close to the spatially averaged value $\tilde{\beta} = 0.46$ for a thermal distribution (at a temperature of one half of the trap depth). This is at least 50 times better than the current systems involving nanofibers ($\beta = 10^{-2}$) \cite{Corzo2019} and in the same range of current PCW-based platforms ($\beta = 0.45$) \cite{Goban2015}.  

Other waveguides combining chiral and strong coupling to emitters exist, however with a common hurdle in maximizing and overlapping spatially both features~\cite{Siampour2023,Lang2022}. In our case, this overlap is readily excellent, due to the absence of symmetry protection from chirality and to the limited spatial structure of the mode [see Figure~4(b)], from which both chirality and Purcell factor inherit. In other words, with our design, we recover a behavior akin to the case of a nanofiber, with the increased group index enhancing the coupling.
 
\section{Trapping Rubidium atoms near a half-W1 waveguide} \label{sec:trap}

In the previous section, simulations were performed for atoms at 116 nm from the edge of the waveguide. Indeed, in the following we show a stable trapping scheme based on an evanescent two-color dipole trap formed by fast guided modes, allowing the atoms to be trapped as close as 115 nm from the surface. This trap has been designed following the ideas implemented in optical nanofibers \cite{LeKien2004,Vetsch2010}, with blue- and red-detuned counter-propagating modes. Finding a stable trapping scheme that keeps the atoms close enough to the surface so that they can couple to the slow mode with a large Purcell factor is a critical requirement for experimental implementations.

\subsection{Two-color dipole trap structure}

In contrast with optical nanofibers, the guided modes in the half-W1 waveguide are structured along the propagation direction due to the Bloch wave structure of the light field. The intensity of the modes, which is an important quantity when looking at dipole trapping, is periodic with period $a$, as shown in figure~\ref{fig:mode}(a) for the slow mode. This feature constrains the position of the trapped atoms to the maxima of intensity of the red-detuned mode. It makes the search for a blue detuned mode more challenging as this one will also be structured, while a uniform one would work perfectly well to repel the atoms from the surface \cite{Goban2012}. A blue-detuned beam with an intensity pattern completely out of phase with the red-detuned one is needed. Fortunately, modes separated by a band gap usually have intensity maxima shifted by $a/2$ \cite{Johnson2004}. We then use the highest available guided band for the blue-detuned trap between 400 and 420~THz (figure~\ref{fig:structure}(c)).

In order to have a full description of the potential seen by the atoms, we take into account the Casimir-Polder (CP) interaction \cite{Casimir1948} between the atoms and the surface. Vacuum fluctuations can polarize the atoms, even if they are not charged. When put in proximity to structures, the vacuum-induced dipole moment creates a mirror charge that acts on the original dipole, leading to an additional light shift. The CP potential $U_{\mathrm{CP}}$ is only significant at very close distances ($\leq 150$~nm) but is crucial as it acts as an attractive potential close to the surface. For these systems, the approximation of an atom in the proximity of an infinite dielectric half space is often used $U^{\textnormal{plane}}_{\textrm{CP}} = - C_3/d^3$ \cite{Johnson2004}, where $d$ is the distance to the surface. As a slab, our structure deviates significantly from a half space. Hence, we computed a more realistic, space-dependent CP potential, based on the pairwise summation technique (PWS) \cite{Bitbol2013}, as described in Appendix \ref{appendix:CP}.

\subsection{Trapping potential simulation}

\begin{figure}
  \includegraphics[width = 0.5\textwidth]{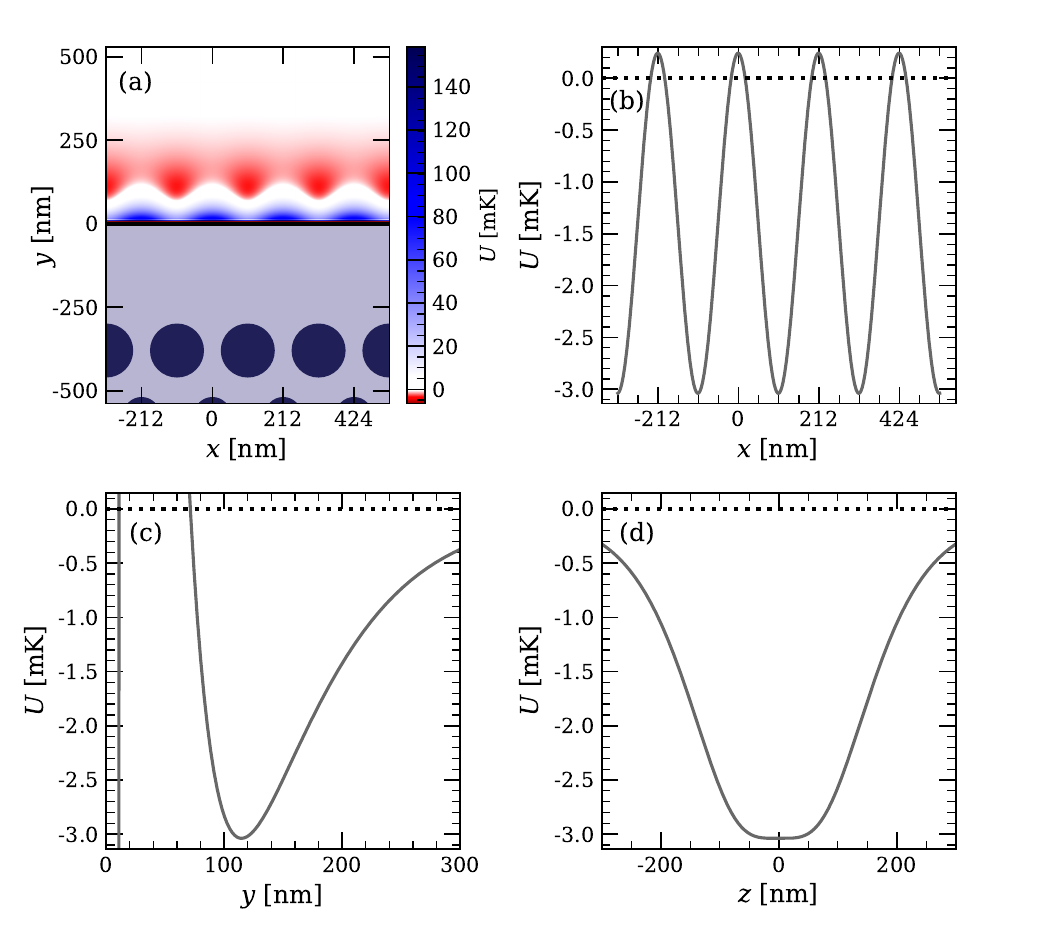}
  \caption{Calculated potential of the two-color dipole trap. (a) 2D total trapping potential $U_\mathrm{tot}$ in the proximity of the waveguide, in the ($x$,$y$) plane. The trapping potential is given along (b) $x$, (c) $y$ and (d) the azimuthal direction out of the symmetry axis. The inset shows the curved trap in the ($y$,$z$) plane. The trap is taken along the dotted white line. A periodic stable trap with depth of about 2.6~mK is obtained with powers of 3.1~mW for each blue beam and 93~$\mu$W for each red one. The simulations are performed with the \texttt{nanotrappy} package.}
  \label{fig:trap}
 \end{figure}

The trapping potentials were obtained via \texttt{nanotrappy}~\cite{Berroir2022}, a Python package developed by our group, to design, calculate and optimize dipole traps around nanoscale waveguides, making the search process faster and more systematic. 

Figure~\ref{fig:trap} shows the total trapping potential $U_\mathrm{tot}$ in 3 directions for an atom in the \mbox{$|F=2, m_F= + 2\rangle$} hyperfine level. $U_\mathrm{tot}$ is the sum of the contributions of the blue potential $U_\mathrm{blue}$, the red potential $U_\mathrm{red}$ and the CP potential $U_\mathrm{CP}$. 
Figures~\ref{fig:trap}(a-c) show the trap along the $x$ and $y$ axis, in the symmetry plane of the waveguide. As shown in figure~\ref{fig:trap}(d), trapping out of this plane is less obvious as the confinement of the atom is low in the azimuthal direction. The curvilinear coordinate is computed as the total distance the atom can travel in the valley of minimal potential shown in the inset with the origin corresponding to $z=0$. The real depth of the trap is given by the potential barriers that appear in figure~\ref{fig:trap}(d), and are physically around $y$~=~0. The trap in this direction is about 100 $\mu$K deep. The low trapping in this direction comes from balancing opposing requirements needed for trapping. Indeed, a stable trap position with a repulsive barrier in the $y$ direction, requires the blue evanescent mode to decay faster than the red one. This criterion, once met, also manifests itself in the azimuthal direction which leads to a predominant attractive trapping almost up to the surface, and the formation of a relatively flat valley in this direction. Even if the atoms can be spread over this valley, the averaged beta factor $\tilde{\beta}$ for a thermal cloud in this trap is still close to the peak value (see above).

For this trap, a beam red-detuned from the D${_2}$ line of $^{87}$Rb at 784.45 nm and a beam blue-detuned at 735.86 nm are used. For each color, another beam detuned by respectively 280 and 385 GHz at the same power is counterpropagated for vector shift cancellation. An absolute trap depth (relative to the atoms being infinitely far away) of 2.6~mK is obtained with a minimum at 116 nm from the surface. The total powers are $P_\mathrm{blue} = 2 \times 3.1$~mW and $P_\mathrm{red} = 2 \times 93~\mu$W, but a stable trap can be obtained over a wide range of powers, with trapping positions ranging typically from 115 to a few hundreds of nm. The main limitation can be the power handling of the structure which is still to be determined. In \cite{Combrie2009}, power densities up to 1 GW/cm$^2$ were coupled to similar GaInP PCWs with group index 8.8. For our structure which has a cross section 10 times smaller and a group index 3 times bigger, this would be equivalent to coupling $\simeq 100$~mW into our waveguide. The proposed powers for the trap fall well below this bound. 

The trapping frequencies are large in the $x$ and $y$ directions, with \mbox{$\omega_x = 2 \pi \times 1.75$~MHz} and \mbox{$\omega_y = 2 \pi \times 2.00$~MHz}. Out of the symmetry plane however, there is an important anharmonicity of the trap in the azimuthal direction. The trap is less constrained in that direction and we extract \mbox{$\omega_z= 2 \pi \times 83$~kHz} from the curvature at the bottom of the trap, which is to be taken with caution given the unusual shape of the trap. 
The addition of a blue-detuned beam with crossed polarization could be used to mitigate this problem, as extensively used in the context of nanofiber traps \cite{Vetsch2010}.

Importantly, we also verified that we can achieve a stable trap in the three directions for a wide range of wavelengths, which is a valuable feature for finding the right trade-off between heating the atoms with off-resonant scattering and power handling of the waveguide. If we allow the blue power to go up to 3~mW, we can find a stable trap for blue wavelengths ranging from 724~nm to 738~nm. Pushing the maximum allowed power to 5~mW we can find a trap for the full available blue-detuned air band, i.e. $\Delta \lambda = 21~\mathrm{nm}$. For the red-detuned laser, we have an available range from 780.5~nm up to 786~nm. Laser diodes are easily available on these wavelengths, reinforcing the feasibility of our platform.

Finally, as briefly noted before, we used counter-propagating beams here instead of single beams, albeit standing waves are not needed for periodic intensity modulation. The strong ellipticity of the guided modes, as shown in figure~\ref{fig:mode}(c), acts as a fictitious magnetic field on the atoms, splitting the Zeeman levels \cite{Cohen1972}. If we start from atoms evenly distributed in all the $m_F$ states, this effect would lead to a large inhomogeneous broadening up to a few GHz. It can be mitigated by using counter-propagating trapping beams slightly detuned from each other, as used for blue detuned beams in some compensated nanofiber traps \cite{Goban2012}. Via \texttt{nanotrappy}, we estimated that adding a red-detuned laser at 280 GHz from the first one and a blue detuned at 385 GHz from the other reduces this broadening by 90$\%$. Counterpropagation creates a running wave at a velocity given by $\delta \omega/k$. This pattern propagates but at a speed so large the atoms only see the average of the potential.

\section{Conclusion} \label{sec:conclu}

Many experimental and technological challenges have yet to be overcome to enable further neutral-atom waveguide-QED protocols. As such, experimental robustness of the targeted waveguide platforms is a critical requirement, as is evanescent trapping of atoms. In our work, we proposed and engineered a bona fide platform for trapping cold Rubidium atoms close to a half-W1 photonic crystal waveguide based on high-index material GaInP. Atoms can be trapped between 115 and a few hundreds of nm from the surface at low input power compatible with the nanophotonic device. At 116~nm, the slow mode couples to the atoms with a Purcell factor of 0.71 with a group index around 28. This study has been carried out for conservative parameters and a strong focus on robustness against fabrication imperfections has been done by engineering the band structure for a large bandwidth, facilitating first implementations. Future generations should support higher group index, albeit with narrower bandwidths~\cite{Li2008}.
This novel platform -- tailor-designed for atom integration, chiral coupling, robustness, large optical access -- offers unique advantages for studying coherent and dissipative dynamics in the waveguide-QED framework. 

\section{Acknowledgements}
This work was supported by the French National Research Agency (NanoStrong Project ANR-18-CE47-0008), by the R\'egion Ile-de-France (DIM SIRTEQ), and by the European Union's Horizon research and innovation program under Grant Agreement No.899275 (DAALI project) and under Grant Agreement No.101097755 (ERC NanoAtom). A.U. was supported by the European Union (Marie Curie Fellowship SinglePass 101030421). J.L. is a member of the Institut Universitaire de France.

\appendix

\section{Theoretical framework: Reaching strong Purcell factor}\label{appendixA}

We can define the spontaneous decay rate of a multilevel atom $\Gamma_{F',F,m_F,q}$ from the hyperfine level \mbox{$|F', m_F - q\rangle$} to \mbox{$|F, m_F\rangle$} as follows \cite{Luan2020}:

\begin{multline}\label{eq:decay}
\Gamma_{F',F,m_F,q} = \frac{2 \mu_0 \lvert\mel{F}{|\hat{\text{d}}|}{F}\rvert^2}{\hbar} \omega_q^2 \lvert C_{m_F,q}\rvert^2 \times \\
 \Im\left[ \hat{\mathbf{e}}_q^* \cdot \mathbf{G}(\mathbf{r},\mathbf{r};\omega) \hat{\mathbf{e}}_q \right]
\end{multline}
where the $\hat{\mathbf{e}}_q$, $q \in \{-1, 0,1\}$, are the normalized dipole vectors over all the possible decay channels ($\sigma^+$, $\pi$, $\sigma^-$ respectively) and $\omega_q$ is the transition frequency between the specified levels. $\mathbf{G}(\mathbf{r}, \mathbf{r}; \omega_q)$ is the value of the classical Green's tensor at the atom position $\mathbf{r}$. The C$_{m_F,q}$ are the Clebsch-Gordan coefficients given by:
\begin{equation}
    C_{m_F,q} = (-1)^{F'-1+m_F} \sqrt{2F+1} \wthreej{F'}{1}{F}{m_F-q}{q}{-m_F}.
\end{equation}

As the modes can be decomposed into an orthogonal basis (with guided and radiative modes forming independent subspaces) \cite{Snyder2012}, it is possible to write the Green's tensor as a sum $\mathbf{G}(\mathbf{r}, \mathbf{r}; \omega_q) = \mathbf{G}_{\text{1D}}(\mathbf{r}, \mathbf{r}; \omega_q) + \mathbf{G}'(\mathbf{r}, \mathbf{r}; \omega_q)$, with $\mathbf{G}_{\textnormal{1D}}$ the part of the Green's tensor corresponding only to the guided mode of interest. From the theory of periodic waveguides \cite{Lecamp2007}, we can derive an analytical expression for the imaginary part of $\mathbf{G}_{\textnormal{1D}}$ of a periodic waveguide at a given position $\mathbf{r}$:
\begin{multline}
\Im\left[ \hat{\mathbf{e}}_q^* \cdot \mathbf{G}_{\textnormal{1D}}(\mathbf{r},\mathbf{r};\omega) \hat{\mathbf{e}}_q \right] = \frac{a}{4 \mu_0 \omega v_\mathrm{g}} \times \\
\left[ \frac{\lvert  \hat{\mathbf{e}}_q \cdot \mathbf{E}^+(\mathbf{r}) \rvert^2}{\int_{\mathcal{V}_{\mathrm{cell}}} d\mathbf{r'} \epsilon(\mathbf{r'})\lvert \mathbf{E}^+ (\mathbf{r'})\rvert^2} + \frac{\lvert  \hat{\mathbf{e}}_q \cdot \mathbf{E}^-(\mathbf{r}) \rvert^2}{\int_{\mathcal{V}_{\mathrm{cell}}} d\mathbf{r'} \epsilon(\mathbf{r'})\lvert \mathbf{E}^+(\mathbf{r'})\rvert^2} \right]
\end{multline}
where $\mathbf{E}^+(\mathbf{r})$ and $\mathbf{E}^-(\mathbf{r})$ are the electric fields of the guided mode in the forward and backward propagating direction respectively, $a$ the period of the modulation, $\epsilon(\mathbf{r})$ the dielectric function of the structure and $v_\mathrm{g} = \pdv{\omega}{k}$ the group velocity of the guided mode. The integrals run over the volume of a waveguide unit cell $\mathcal{V}_{\mathrm{cell}}$ of length $a$.

In the absence of a external magnetic field (Zeeman degeneracy), the 1D decay rate of a single atom from \mbox{$|F', m_F - q\rangle$} to \mbox{$|F, m_F\rangle$} through a single decay channel and into the guided mode of interest of the waveguide is hence given by:
\begin{multline}
\Gamma_{\mathrm{1D},F',F,m_F,q} = \frac{\pi a c}{\hbar}
\frac{\lvert\mel{F}{|\hat{\mathrm{d}}|}{F}\rvert^2}{\lambda_0 v_\mathrm{g}}
\lvert C_{m_F,q}\rvert^2 \times \\
\left[\frac{\lvert  \hat{\mathbf{e}}_q \cdot \mathbf{E}^+(\mathbf{r}) \rvert^2}{\int_{\mathcal{V}_{\mathrm{cell}}} d\mathbf{r'} \epsilon(\mathbf{r'})\lvert \mathbf{E}^+ (\mathbf{r'})\rvert^2} + \frac{\lvert  \hat{\mathbf{e}}_q \cdot \mathbf{E}^-(\mathbf{r}) \rvert^2}{\int_{\mathcal{V}_{\mathrm{cell}}} d\mathbf{r'} \epsilon(\mathbf{r'})\lvert \mathbf{E}^+ (\mathbf{r'})\rvert^2}\right] \label{eq:Gamma1D_multilevel}
\end{multline}

The 1D Purcell factor $\Gamma_{\mathrm{1D}}/\Gamma_0$ which is often used to quantify the decay of the atoms to the guided mode can be obtained from equation \eqref{eq:Gamma1D_multilevel} by dividing it by the single atom free space decay rate $\Gamma_0$.

The first (second) term in \eqref{eq:Gamma1D_multilevel} corresponds to the emission of the atom into the guided mode in the forward (backward) direction. By reciprocity, we can define the excitation rates $\gamma_{\mathrm{exc},F,F',m_F,q}^{\pm}$ of an atom initially in \mbox{$|F, m_F\rangle$} and promoted to \mbox{$|F', m_F - q\rangle$} when coupled to an input propagating mode at the resonance frequency:
\begin{multline}
\gamma_{\mathrm{exc},F,F',m_\text{F},q}^\pm = \frac{\pi a c}{\hbar}
\frac{\lvert\mel{F}{|\hat{\text{d}}|}{F}\rvert^2}{\lambda_0 v_\mathrm{g}}
\lvert C_{m_F,q}\rvert^2 \times \\
\frac{\lvert  \hat{\mathbf{e}}_q \cdot \mathbf{E}^\pm(\mathbf{r}) \rvert^2}{\int_{\mathcal{V}_{\mathrm{cell}}} d\mathbf{r'} \epsilon(\mathbf{r'})\lvert \mathbf{E}^+ (\mathbf{r'})\rvert^2} \label{eqn:exc}
\end{multline}
where $\mathbf{E}^\pm(\mathbf{r})$ is here the mode of the excitation laser injected in the waveguide (forward or backward propagating). We note that $\Gamma_{\mathrm{1D},F',F,m_F,q} = \gamma_{\mathrm{exc},F,F',m_F,q}^{+} + \gamma_{\mathrm{exc},F,F',m_F,q}^{-}$.\\

As such, we see that to reach high excitation rates and Purcell factors we must decrease $v_\mathrm{g}$, which can be achieved by dispersion design, and maximize the normalized electric field amplitude at the position of the atom given by the second half of equation~\eqref{eqn:exc}. 

When considering this time the decay rate (equation \eqref{eq:Gamma1D_multilevel}) along a circular dipole transition $\sigma^\pm$, usually only one of the two terms is dominant because of the strong chirality of the modes of the waveguide. For a $\pi$ transition however, the dipole vector is real and since $\mathbf{E}^+(\mathbf{r})^*$~=~$\mathbf{E}^-(\mathbf{r})$ for non-absorbing materials, the two terms of the sum are equal $\gamma_{\mathrm{exc},F',F,m_F,0}^+ = \gamma_{\mathrm{exc},F',F,m_F,0}^-$.

\section{Casimir-Polder interactions between GaInP and Rubidium atoms}\label{appendix:CP}

\begin{figure}
\centering
  \includegraphics[width = 0.5\textwidth]{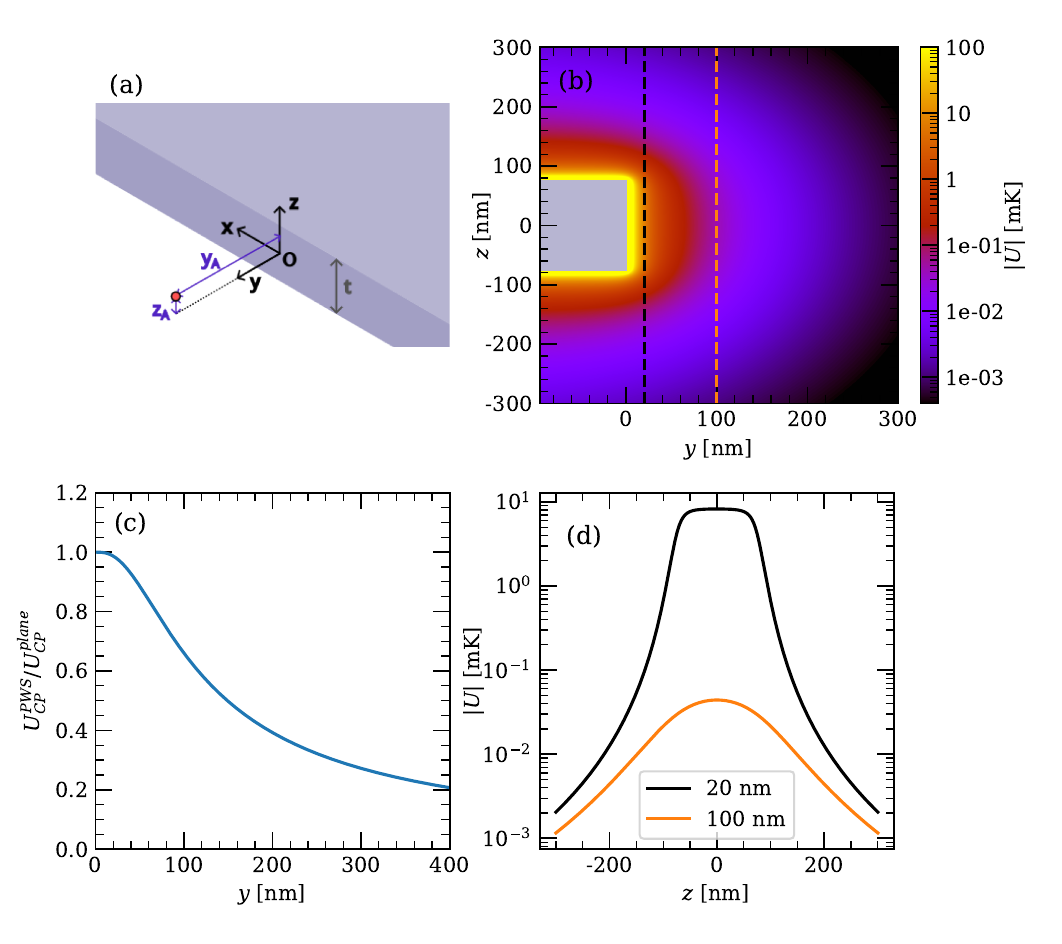}
  \caption{Calculation of the Casimir-Polder interaction between the structure and the atom with PWS. (a) Coordinate system used with the simplified slab of thickness $t$ (b) 2D log plot of the magnitude of the CP in the transverse plane of the crystal (c) Comparison between the PWS potential and the one for an semi-infinite dielectric plane. As expected, they are comparable at small distances, as the solid angle under which the atom interacts with the surface becomes large. (d) Shape of the vertical dependence of the interactions at different distances from the surface. At close distances we have a plateau and fast decay on each side. At 100~nm the shape is very different with a slower decay.} \label{fig:CP}
 \end{figure}

In order to get a full description of our trap, we include the Casimir-Polder (CP) interactions as they are not negligible at such close distances from the surface. A crude approximation would be that the atom is close enough to the structure such that the well-known formula for the CP potential for an atom close to a semi-infinite dielectric wall can be used $U_\textnormal{CP}^{\textnormal{plane}} = - C_3/d^3$ \cite{LeKien2004, Stern2011}. However, atoms in our trap explore way beyond this approximation, both with the distance to the surface being comparable to the slab thickness, as well as from the azimuthal extension. Assuming an infinite surface would create a fake barrier at $y=0$ and $z>t/2$ whereas using the formula but limiting it to the range $-t/2<z<t/2$ would create a discontinuity at these lines.\\

A formula based on scattering theory can be used for computing these CP interactions \cite{Lambrecht2006}. If analytical solutions have only been found for simple geometries \cite{Antezza2004}, it is possible to write it in terms of the scattering Green's tensor and use an electromagnetic solver to compute the latter \cite{Rodriguez2009}. As this comes at the expense of very intensive computations, we turn to an approximated and simpler derivation of these interactions from first principles. 
We sum the van der Waals (vdW) interaction between the trapped atom and all the atoms constituting the structure. This method, referred to as pairwise summation (PWS), assumes the vdW potentials are additive, neglecting collective effects in the material. The magnitude of these collective effects can be significant, yet it has been shown that the results are within $30\%$ of the exact calculation \cite{Sparnaay1959, Bitbol2013} which is sufficient for our purposes.\\

Hence, we write : 

\begin{equation}
 U^{PWS}_\textnormal{CP} = \int_{\mathcal{V}} U_{\textnormal{vdw}},
\end{equation}
where we will consider $U_{\textnormal{vdw}}$ as $A/r^6$, the approximation for short range interactions. $A$ is a constant that depends on the polarizabilities of the atoms involved in the integral~\cite{Bitbol2013}. The constants are not crucial here as they will be determined afterwards by comparing our result to the known one for an infinite plane. The integral runs over the whole volume $\mathcal{V}$ of the waveguide.
For simplicity we assume the waveguide to be a semi-infinite slab of thickness $t$ (see figure~\ref{fig:CP}(a)). This allows us to write $r = \sqrt{x^2 + (y'-y_a)^2 + (z'-z_a)^2}$, where ($y_a$,$z_a$) denote the position of the atom. $x_a$ is irrelevant because the structure is infinite in this direction.

\begin{multline}
U^{PWS}_\textnormal{CP} = A \int_0^{-\infty} dy \int_{-t/2}^{t/2} dz \int_{-\infty}^{+\infty} dx  \times \\ 
\frac{1}{(x^2 + (y-y_a)^2 + (z-z_a)^2)^3}
\end{multline}

The integrals can be calculated analytically and we obtain: \\

\begin{widetext}
\begin{multline}
U_\textnormal{CP}(y_a,z_a) = - \frac{A \pi}{8}\left[\frac{2y_a^3 - \sqrt{1/(y_a^2+(t-z_a)^2}(2y_a^4+y_a^2(t-z_a)^2+2(t-z_a)^4}{3y^3(t-z_a)^3} \right. \\
+ \left. \frac{2y_a^3 - \sqrt{1/(y_a^2+(t+z_a)^2}(2y_a^4+y_a^2(t+z_a)^2+2(t+z_a)^4}{3y_a^3(t+z_a)^3} \right] \label{eq:CP_PWS}
\end{multline}
\end{widetext}

We note that when $t \to +\infty$, equation~\eqref{eq:CP_PWS} reduces to $U_\textnormal{CP} = \frac{A \pi}{6 y_a^3}$, and we recover the usual dependence for an atom in front of a dielectric half-space. Comparing this to $U_\textnormal{CP}^{\textnormal{plane}}$ we find $A = \frac{6C_3}{\pi}$.
 
The obtained spatial dependence of the CP interactions is shown in figure~\ref{fig:CP}(b). Figure~\ref{fig:CP}(c) shows that with this expression we depart significantly from the infinite plane formula after only 100~nm as the finite thickness of the slab cannot be neglected. Finally, figure~\ref{fig:CP}(d) shows how the transverse shape of the potential changes with distance, an important feature for our trapping scheme, which is not encapsulated in $U_\textnormal{CP}^{\textnormal{plane}}$. 
\\

We then have to determine the value of the constant $C_3$ in order to have a full description of the potential. To the best of our knowledge, there were no previous computations of the $C_3$ coefficient of the CP interactions between GaInP and Rubidium atoms. For a dielectric wall, we use the following formula for $C_3$ \cite{Caride2005}:
\begin{equation}
C_3 \approx \frac{\hbar}{4\pi} \int^{+\infty}_0 \alpha(i\xi)\frac{\epsilon(i\xi) - 1}{\epsilon(i\xi) + 1} d\xi.
\label{c3}
\end{equation}
where $\alpha$ is the Rubidium scalar polarizability and $\epsilon$ the dielectric constant of GaInP.
This formula requires evaluating $\alpha$ and $\epsilon$ over the imaginary axis. $\alpha(i\xi)$ is directly evaluated using the expression for the scalar polarizability of $^{87}$Rb in the ground state \mbox{$|\mathrm{F}=2\rangle$} with complex frequencies.

As $\epsilon(\omega) = \epsilon'(\omega) + i \epsilon''(\omega)$, we can get the dependence in $\omega$ from experimental data~\cite{Schubert2005}. To evaluate it over the imaginary axis we use the Kramers-Kroenig relation that relates this function over the imaginary axis to $\epsilon'(\omega)$ \cite{Antezza2004}
\begin{equation}
\epsilon(i\xi) = 1 + \frac{2}{\pi} \int_0^{+\infty} \frac{\omega [\epsilon'(\omega)-1]}{\omega^2+\xi^2}d\omega.
\end{equation}
Finally, for a ground state Rubidium 87 atom close to a GaInP surface we get $C_3~=~9.25 \times 10^{-49} \mathrm{J.m^3} = \mathrm{h}\times~1391~ \; \mathrm{Hz.\mu m^{3}}$.

\bibliography{HW1Proposal.bib}
\end{document}